\documentclass{article}
\usepackage{amssymb}
\usepackage{amsmath}
\usepackage{epsfig}
\usepackage{here}
\usepackage{a4}

\begin{document}

\vspace*{1cm}
\begin{center}
{\Large{\bf  POSITRONIUM AS A PROBE FOR NEW PHYSICS BEYOND THE STANDARD MODEL\footnote{Talk
given at the Workshop on Positronium Physics, Zurich (Switzerland), 30-31 May 2003.}}}\\
\vspace{.5cm}
{\large A. Rubbia}\footnote{Andre.Rubbia@cern.ch}

Institut f\"{u}r Teilchenphysik, ETHZ, CH-8093 Z\"{u}rich,
Switzerland
\end{center}
\vspace{2.cm}

\begin{abstract}
We discuss and summarize some aspects concerning the study of 
the positronium system to probe new
physics beyond the Standard Model.
\end{abstract}


%%%%%%%%%%%%%%%%%%%%%%%%%%%%%%%%%%%%%%%%%%%%%%%%%%%%%%%%%%%%
% The main text of your paper   begins here                          %
%%%%%%%%%%%%%%%%%%%%%%%%%%%%%%%%%%%%%%%%%%%%%%%%%%%%%%%%%%%%
\newpage
\section{Introduction}

Quantum electrodynamics (QED) is the textbook example of the success of quantum field theory.
Many physical quantities (anomalous electron and muon magnetic moments, hyperfine splitting 
of hydrogen, muonium and positronium, Lamb shift, etc.) have been calculated very precisely. The
measurements have been characterized by excellent agreement with the theoretical predictions and
in general provided very clean conditions for hunting for small deviations from the standard theory.

Positronium ($Ps$), the positron-electron bound state,
 is the lightest known atom, which is bounded and self-annihilates 
through the same, electromagnetic interaction. At the 
current level of experimental and theoretical precision this is  
the only interaction present in this system. 
This feature has made positronium an ideal system for  
testing the accuracy of QED calculations 
for bound states, in particular for the triplet ($1^3S_1$)
state of $Ps$, orthopositronium ($o-Ps$). 
Due to the odd-parity under
C-transformation  $o-Ps$ decays
predominantly into three photons. 
 As compared with the singlet ($1^1S_0$) state (parapositronium),
 the "slowness" of $o-Ps$ decay rate, due to the phase-space and  additional
$\alpha$ suppression factors, gives an enhancement factor $\simeq 10^3$,
making it more sensitive to an admixture of 
new interactions which are not accommodated in the Standard Model.

Positronium was discovered experimentally by Deutsch\cite{Deutsch} in 1951, who
observed its decay in different gases. Since then, a lot of focus has been set on the determination
of its basic properties like decay life time, decay modes, spectroscopy, etc. In particular, the
measurement of the  o-Ps lifetime caught much attention.

Since the 1989,
the precision on the o-Ps lifetime reached a value well under 1000 ppm. Much excitement
arose when the measurements performed by the Michigan group did not agree
with theory. This problem, called the o-Ps-lifetime-puzzle, ignited much experimental and
theoretical activity devoted to its clarification. These are: (1) new direct lifetime measurements
by the Tokyo group which did not confirm the discrepancy (2) new theoretical calculations by Adkins et al. 
including higher order terms improving the theoretical precision well below experimental
errors, however, confirming early theoretical estimates (3) searches for ``exotic" decay modes
which could explain the lifetime discrepancy at the cost of new physics (violation of
basic conservation laws with decays into 1 photon,
2 photons; anomalous rate in 5 photons; millicharged particles; new bosons, ...)   (4) exotic
suggestions for disappearance mechanisms (mirror worlds, extra dimensions). So far
none of these provided a clear solution to the o-Ps-lifetime-puzzle.

Just before the Positronium physics workshop held in Zurich in May 2003, 
the Michigan group has published a new result
which is now in agreement with the theoretical value, somewhat in contradiction
with the earlier results from the same group. The easy shortcut is to assume
that the o-Ps-lifetime puzzle is closed. Should we however indeed consider
that the o-Ps lifetime puzzle is solved and hence assume that 
further precise measurements of positronium (o-Ps lifetime, ...) are irrelevant?

During the workshop we tried to obtain a modern view on this problem, in particular
in the context of the physics of positronium.  More generally and in addition
to the puzzle of the orthopositronium lifetime, we tried to address
the study of positronium as a probe for new physics beyond the Standard
Model. In this paper, we review some of these aspects. 

In particular, we discuss the ideas for an ``appearance'' of an effect, namely
through the search for the invisible decay of the o-Ps, i.e. a photon-less decay.   
It may be worthwhile to remember that the process with analogous experimental signature, 
$Z \rightarrow invisible$ decay 
plays a fundamental role in the determination of the number of lepton families.  
This is to be contrasted with
the ``disappearance'' searches, where one measures the o-Ps-lifetime as precisely as possible
and then looks for a deviation of this result from the theoretical one. In the disappearance
mode, one is looking for a small effect. In the appearance mode, we are looking for
few events with energy deposition in our detector compatible with zero, a direct experimental signature that
cannot happen if o-Ps decays to standard particles. New experiments are being
designed (see Section~\ref{sec:invis}) to reach a sensitivity in the branching ratio at the level of $10^{-8}$.

In addition, we mention the fact that a new experiment capable for investigating the invisible
o-Ps decay down to $10^{-8}$ could test the existence of extra-dimensions.
This could turn out to be a new domain of research and stresses once again
the known fact that precise experiments at low energy can compete with big projects at the
highest energy frontiers. 
Note that within the Standard electroweak model,  orthopositronium  can  decay invisibly into a
neutrino-antineutrino pair.   The $o-Ps \to \nu_e \bar{\nu}_e$ decay occurs through 
$W$ exchange in the $t$--channel and $e^+e^-$ annihilation via $Z$.
The decay width is \cite{czar}
\begin{eqnarray} 
\Gamma(o-Ps \to \nu_e  \bar{\nu}_e) 
%\frac{G2_F \alpha^3m2_e}{24\pi2}(1 + 4 \sin2\theta_W)2 
\approx 6.2 \cdot 10^{-18}\Gamma_{3\gamma}
\end{eqnarray}
For other neutrino flavours only the $Z$-diagram 
contributes. For $ l \neq e$ the decay width is \cite{czar}
\begin{eqnarray} 
\Gamma(o-Ps \rightarrow \nu_l \bar{\nu}_l) 
%\frac{G2_F \alpha^3m2_e}{24\pi2}(1 - 4 \sin2\theta_W)2 
\approx 9.5 \cdot 10^{-21}\Gamma_{3\gamma}
\end{eqnarray}
Thus, in the SM the $o-Ps \to \nu \bar{\nu}$ decay rate
is very small and evidence for invisible decays  
 would unambiguously signal the presence of new physics.  

Overall, precise study of positronium could yield
new information on fundamental discrete symmetries like charge conjugation (C), parity (P),
time-reversal (T), CP and even CPT, continuing the long tradition of tests
of these fundamental symmetries in atomic systems.

Of course, the study of positronium cannot be accomplished without a precise
theoretical understanding of the QED predictions. We also collected for the workshop
theoretical contributions in the context of precision tests of the QED theory,
high order QED corrections, connection between positronium and quarkonium.

During this workshop we also wanted to have a critical review of the existing
o-Ps-lifetime results and planned some talks accordingly, which resulted in interesting
discussions during the meeting.

Finally, one aspect that has been neglected in this introduction up to now
is the possible application of the experimental techniques developed in the
context of fundamental positronium physics to the applied science of materials.
To address this point, we have invited two talks related to the use of positron annihilation
for the characterization of solids and polymers.

\section{Renaissance of the mirror world ?}
In 1956 Lee and Yang proposed\cite{Lee:qn}  that the weak interactions of fundamental
particles were not invariant under the parity transformation. They suggested it to explain
some experimental results that were considered as puzzles. They pointed out that
these puzzles could be explained if one assumed a left-right asymmetry in the weak
interactions. In their original paper, they also provided a ``remedy" for this a priori 
not sensible assumption. 
Indeed Lee and Yang argued\cite{Lee:qn} that 
if left-right asymmetry were found in weak interactions, the question could still be raised whether there could not exist corresponding elementary particles exhibiting opposite asymmetry such that in the broader sense there will still be over-all right-left symmetry. They spoke
specifically of two kinds of protons, the left-handed and the right-handed one. As well known, left-handed nature of ordinary
matter was brilliantly confirmed experimentally.
Nowadays parity violation in fundamental interactions is so well accepted, 
that the left-right asymmetry of Nature is inserted in the modern Standard Model
``from the beginning" in the assignment of the particle fields. Nonetheless, the question
whether Nature is fundamentally left-right symmetric or not has remained up to now unresolved.

Landau was always convinced by the absolute symmetry of vacuum\cite{Okun:2001kg}. Under
his appeal Kobzarev, Okun and Pomeranchuk suggested the hypothesis of a mirror world\cite{flv}.

It is fair to say that the existence of mirror matter has been recently boosted
since the realization of its possible connection to the dark matter problem. If about 30\% of
the critical density of the Universe is composed of non-baryonic dark matter, as
recent astrophysical observations seem to indicate, then this clearly motivates
the search for new kinds of matter, since the current Standard Model
does not contain any heavy, stable non-baryonic particles.

The idea that there can exist a hidden mirror sector of
particles and interactions which is the exact duplicate
of our visible world 
has attracted a significant interest over last years and
has been summarized at this workshop by Berezhiani\cite{berezhiani}. 
The basic concept is to have a theory given by the
product $G\times G'$ of two identical gauge factors with
the identical particle contents, which could naturally emerge
e.g. in the context of $E_8\times E'_8$ superstring. 

As discussed by Foot\cite{foot}, the ordinary and mirror particles could form parallel sectors each
with gauge symmetry $G$ (where $G=G_{SM} \equiv SU(3)_c \otimes SU(2)_L 
\otimes U(1)_Y$
in the simplest case) so that the full gauge group is $G \otimes G$.
Mathematically, mirror symmetry has the form:\cite{flv}
\begin{eqnarray}
& x \to -x, \ t \to t, \nonumber \\
& W^{\mu} \leftrightarrow W'_{\mu}, \ B^{\mu} \leftrightarrow B'_{\mu},
\ G^{\mu} \leftrightarrow G'_{\mu} \nonumber \\
& \ell_{iL} \leftrightarrow \gamma_0 \ell'_{iR}, \
e_{iR} \leftrightarrow \gamma_0 e'_{iL}, \
q_{iL} \leftrightarrow \gamma_0 q'_{iR}, \
u_{iR} \leftrightarrow \gamma_0 u'_{iL}, \
d_{iR} \leftrightarrow \gamma_0 d'_{iL}, 
\end{eqnarray}
where $G^{\mu}, W^{\mu}, B^{\mu}$ are the standard
$G_{SM}$ gauge particles,
$\ell_{iL}, e_{iR}, q_{iL}, u_{iR}, d_{iR}$ are the
standard leptons and quarks ($i=1,2,3$ is the generation index)
and the primes denote the mirror particles. 

Ordinary and mirror particles couple with each other via gravity
and possibly by new interactions connecting ordinary and mirror
particles together. 
Constraints from gauge invariance, mirror symmetry and
renormalizability, suggest only two
types of new interactions\cite{flv}:
a) Higgs-mirror Higgs quartic coupling
(${\cal L} = \lambda' \phi^{'\dagger}\phi' \phi^{\dagger} \phi$),
and b) via photon-mirror photon kinetic mixing:
\begin{eqnarray}
{\cal L}_{int} = {\epsilon \over 2}F^{\mu \nu}F_{\mu \nu}' .
\label{km}
\end{eqnarray}
where $F^{\mu \nu}$ ($F'_{\mu \nu}$)
is the field strength tensor for electromagnetism (mirror
electromagnetism).
The effect of photon-mirror photon kinetic mixing is to
cause mirror charged particles
to couple to ordinary photons with effective electric
charge $\epsilon e$\cite{flv,hol,s}.

\subsection{Mirror world and invisible decays of positronium}
\label{sec:mirrorworld}
\begin{figure}[htb]
\begin{center}
{\epsfig{file=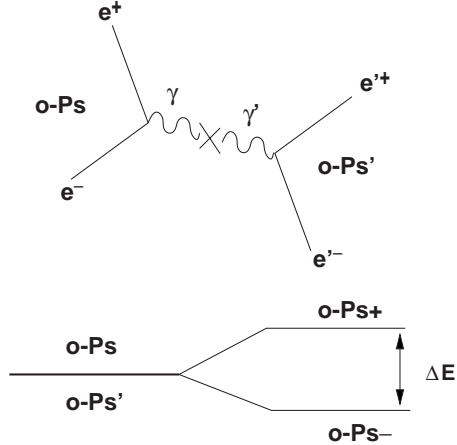,width=60mm}}
\end{center}
 \caption{\em The double degeneracy between orthopositronium mass eigenstates
of ordinary o-Ps and mirror o-Ps' is broken when a small mixing 
(upper picture) term is included. }
\label{mixing}
\end{figure}

Glashow pointed out\cite{gl} that a sensitive laboratory test for mirror
matter comes from the orthopositronium
system. The interaction of $e^+ e^-$ with $e^{'+} e{'-}$
leads to a small mass term mixing orthopositronium with
mirror orthopositronium. 
The effect of this mass mixing
term is to cause orthopositronium to (maximally) oscillate
into mirror orthopositronium\cite{Foot:2000aj}:
\begin{equation}
P(o-Ps \to o-Ps') = \sin^2 \omega t,
\label{eq:osc}
\end{equation}
where $\omega = 2\pi \epsilon f$, where $f = 8.7 \times 10^4$ MHz
is the contribution to the ortho-para splitting from the one photon
annihilation diagram involving orthopositronium (see Figure~\ref{mixing}).

In an experiment, mirror orthopositronium decays are not 
detected, hence leading to $o-Ps\rightarrow invisible$ decays, which means that
the number of orthopositronium, $N$, satisfies\cite{gl}
\begin{eqnarray}
N = \cos^2 \omega t e^{-\Gamma^{3\gamma}t} \approx exp[-t(\Gamma^{3\gamma} +
\omega^2 t)]
\end{eqnarray}
where $\Gamma^{3\gamma}$ is the standard 
orthopositronium decay rate.
Evidently, the observational effect of the oscillations is to
\textit{increase} the apparent decay rate of ordinary orthopositronium:
$\Gamma^{eff} \approx \Gamma^{3\gamma}(1 + \omega^2/\Gamma^{3\gamma})$. This implies
that the existence of mirror matter can be probed with positronium either (1) via
a precision measurement of its decay rate and/or shape of its decay time spectrum or (2) 
via a direct search for invisible decays.

In the simplest case of $o-Ps \to o-Ps'$ 
oscillations in vacuum\cite{gl} the branching ratio occurring during a long
enough observation time can be calculated as   
\begin{equation}
Br(o-Ps\to invisible) = \frac{2(2\pi \epsilon f)^2}{\Gamma^2_{3\gamma} + 
4(2\pi \epsilon f)^2}
\label{br}
\end{equation}
However,  Eq.(\ref{br}) may not be applicable to
all measurements. In experiments, orthopositronium is not produced in vacuum, but rather
by slow positron collisions with a positronium formation target. As a result,
the newly formed positronium will
undergo elastic collisions  with the target at a rate, $\Gamma_{coll}$, which
depends on the particular experiment. 
These collisions cause quantum decoherence, disrupting the ordinary-mirror oscillations.

Hence, it is absolutely fundamental
to distinguish experiments in ``vacuum'', where o-Ps is contained
for example in a large evacuated cavity, from other setups where
o-Ps undergoes numerous interactions with the environment within
its lifetime. This situation is summarized in Table~\ref{tab:gammcoll},
where the three types of experiments (gas\cite{Michigan_1989}, vacuum cavity\cite{Michigan_1990} and powder\cite{Tokyo_1995},
see Section~\ref{sec:opuzzle}) are illustrated.
It is well known that collisions damp the oscillations, e.g. 
 in the limit where the collision rate is much larger
than the decay rate (or oscillation frequency, whichever is smaller)
the effect of the oscillations becomes negligible.
In addition, external fields might result in a
loss of coherence due to additional splitting of mass eigenstates.

\begin{table}[tbh]
\caption{Some measurements of the orhopositronium lifetime. The last column is an estimate 
of the mean collision rate in the experiment (from Ref.\protect\cite{Foot:2000aj}).}
\begin{tabular}{ccccc} 
\hline
 group/ref & Rate  $\Gamma_{3\gamma}$ & errors & Technique& $\Gamma_{coll}$\\
&  & $\mu s ^{-1}$ &   &\\ \hline
 \hphantom{0}Ann Arbor [\cite{Michigan_1989}] &   7.0514    \hphantom{000} &200& Gas&  $\approx 10^3\Gamma_{3\gamma}$ \\
 \hphantom{0}Ann Arbor[\cite{Michigan_1990}] &   7.0482    \hphantom{000} &230& Vacuum&  $\approx (3-10)\Gamma_{3\gamma}$ \\
 \hphantom{0}Tokyo [\cite{Tokyo_1995}]       &   7.0398    \hphantom{000} &412& Powder&  $\approx 10^4\Gamma_{3\gamma}$ 
\label{tab:gammcoll}
\end{tabular}
\end{table}

Let us   consider the case where the collision rate is much larger than the
 decay rate,
$\Gamma_{coll} \gg \Gamma_{3\gamma}$, then the observed
decay rate is approximately given by\cite{Foot:2000aj}:
\begin{equation}
\Gamma_{obs} \simeq \Gamma_{3\gamma} + {2\omega^2 \over \Gamma_{coll}}
= \Gamma_{3\gamma}\left(1 + {2(2\pi \epsilon f)^2 \over \Gamma_{coll}\Gamma_{3\gamma}}\right). 
\end{equation}
Ignoring the result obtained in gas, the discrepancy originally obtained in the Ann Arbor experiment
could be explained if $\epsilon \approx 10^{-7}$.

\subsection{Mirror world and dark matter}
Foot\cite{foot} discussed the possible importance of mirror matter in the context
of the dark matter puzzle: if mirror matter is identified with the dark matter
in the Universe, then it is natural to interpret the dark
matter halo of our galaxy as containing mirror matter (possibly mirror stars/planets/dust
and gas\cite{foot}).  If the dark matter halo
of our galaxy is composed of mirror matter, then it can potentially
be detected in dark matter experiments via the nuclear recoil
signature via interactions between nuclei and mirror-nuclei
induced by the photon-mirror photon kinetic term mixing.

\begin{figure}[htb!]
\begin{center}
\vspace{-0cm}\hspace{-0.cm}{\epsfig{file= 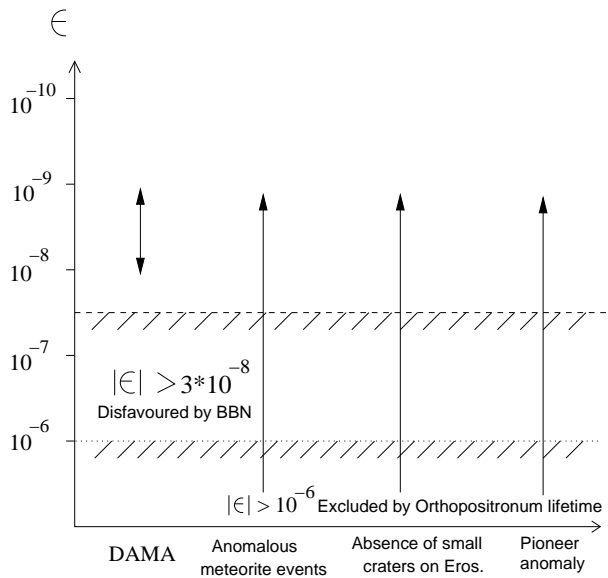,width=80mm}}
\end{center}
\caption{\em Summary of the experimental constraints for $\epsilon$ by Foot\protect\cite{foot}.}
\label{fig:footsum}
\end{figure}

Very strikingly Foot recently pointed out\cite{f03} that the DAMA results\cite{dama2} could
be interpreted in terms of such interactions, if one interprets
the signal in terms of mirror
$O'$, $Fe'$ mixture with an annual modulation effect
in the 2-6 keV window:
\begin{eqnarray}
|\epsilon | \sqrt{{\xi_{O'} \over 0.10} +
{\xi_{Fe'} \over 0.02}} 
\simeq 4.5 \times 10^{-9}
\label{dama55}
\end{eqnarray}
where $\xi_{A'} \equiv \rho_{A'}/(0.3 \ {\rm GeV/cm^3})$ is the $A'$
proportion (by mass) of the halo dark matter.
This is an extremely impressive result which provides an indication for possible values of $\epsilon$ !
These are summarized in Figure~\ref{fig:footsum}
by Foot\cite{foot}. An interesting region, which seems also to be in the range
of naturally small $\epsilon$ motivated by Grand Unification models as discussed
by Berezhiani\cite{berezhiani}, is $10^{-9}\lesssim\epsilon\lesssim  10^{-8}$. Using
Eq.\ref{eq:osc} this implies $10^{-8} < Br(o-Ps\rightarrow invisible) < 10^{-6}$.

\subsection{Cosmology of mirror matter}
How does the picture of mirror matter as dark matter candidate
fit in the global picture of the evolution of the Universe
as predicted by cosmology? Berezhiani\cite{berezhiani} brought up
important points concerning these issues:
naively mirror parity implies that
ordinary and mirror sectors should have the same cosmology, 
and so ordinary and mirror-particles should have the same cosmological 
densities. However, this would be in the immediate conflict 
with the Big Bang nucleosynthesis (BBN) bounds
on  the effective number of extra light neutrinos.
Therefore, the mirror particle density in the early
universe should be appropriately reduced.
This situation is plausible if the following conditions  
are satisfied\cite{berezhiani}:
\begin{itemize}
\item[A.] At the Big Bang the two systems are born with different 
densities.
\item[B.] The two systems interact very weakly,  
so that they do not come into the thermal equilibrium
with each other in the early Universe.
\item[C.] Both systems expand adiabatically and there is no significant 
entropy production at the later times which could heat the 
mirror-sector and equilibrate its temperatures to the ordinary one. 
\end{itemize}

Because of the temperature difference, in the
mirror sector all key epochs as the baryogenesis,
nucleosynthesis, etc. proceed at somewhat different
conditions than in the observable universe.
In particular, in certain baryogenesis scenarios 
the mirror-world generically should get a larger 
baryon asymmetry than the ordinary sector, and  
it is pretty plausible that dark matter of the Universe 
or at least its significant fraction,     
is constituted by mirror-baryons which are obviously dark 
for the ordinary observer\cite{BCV}. 

%%%%%%%%%%%%%%%%%%%%%%%%%%%%%%%%%%%%%%%%%%%%%%%

\section{Extra space-time dimensions and invisible decays of positronium}
Recently  the models with infinite 
additional dimensions  
of the Randall-Sundrum type (brane-world models) have become very popular. 
There is a  hope 
that  models with a big compactification radius 
 will provide the natural solution to the 
gauge hierarchy problem.
For instance, as it has been shown by Randall and Sundrum 
in the five dimensional model,  there exists a 
thin-brane solution to the 5-dimensional Einstein equations which 
has flat 4-dimensional hypersurfaces,
\begin{equation}
ds^2 = a^2(z)\eta_{\mu\nu}dx^{\mu}dx^{\nu} -dz^2.
\end{equation} 
Here
\begin{equation}
a(z) = exp(-k(z-z_c))
\end{equation}
and the parameter $k > 0 $ is determined by the 5-dimensional Planck 
mass and bulk cosmological constant. For the model with 
metric of Randall-Sundrum, the effective four-dimensional gravitational constant is
\begin{equation}
G_{(4)} = G_{(5)}k \frac{1}{exp(2kz_c) - 1}
\end{equation}
One can solve the gauge hierarchy problem in this model if  
$k \sim M_{EW} =  1~TeV$, $G_{(5)}\sim k^{-3}$. It follows
that the Planck scale in this model is
\begin{equation}
M_{PL} \sim exp(kz_c)M_{EW}
\end{equation}
that means the existence of exponential hierarchy between Planck 
and electroweak scales. For $z_c \approx 37 \cdot k^{-1}$ we have 
a correct quantitative relation between Planck and electroweak 
scales.  

In Randall-Sundrum models, physical particles 
are trapped on a three-dimensional brane via curvature in the bulk 
dimension. Although massive particles can indeed be trapped on the brane, 
they are also expected to be metastable\cite{Dubovsky:2000am}. 
That is, the quasi-normal modes are metastable states, that 
can decay into continuum Kaluza-Klein modes in the higher dimensions.
For an observer on the three-brane, massive particles will appear 
to exist for some time and then ``disappear" into the bulk fifth dimension.
   
It was noted\cite{Gninenko:2003nx} that $o-Ps$ is a good candidate 
for searching for this effect of disappearance into 
additional dimension(s) since it has specific quantum numbers similar 
to those of the vacuum and is a system which allows its constituents a rather 
long interaction time.
For the orthopositronium invisible decay into 
additional dimension(s) 
\begin{equation}
o-Ps \rightarrow \gamma^{*} \rightarrow  additional~dimension(s) 
\end{equation}
the corresponding branching ratio is\cite{Gninenko:2003nx}
\begin{eqnarray}
Br(o-Ps \rightarrow \gamma^{*} \rightarrow additional~dimension(s))& 
=\\ \nonumber 
\frac{9\pi}{4(\pi^2-9)}\cdot \frac{1}{\alpha^2}\cdot \frac{\pi}{16}
(\frac{m_{o-Ps}}{k})^2   \approx 3\cdot10^{4}(\frac{m_{o-Ps}}{k})^2
\end{eqnarray}
An important bound on the parameter $k$ arises from data on the $Z\to invisible$ 
decay, which leads to  $k  \gtrsim 2.7~TeV$. Using this, we find  
\begin{equation}
Br(o-Ps \rightarrow additional~dimension(s)) \lesssim 4 \cdot 10^{-9}
\end{equation}
To solve the gauge hierarchy 
problem models with additional infinite dimension(s) 
must have the $k \lesssim O(10)~TeV$. It means that   
\begin{equation}
Br(o-Ps \rightarrow additional~dimension(s)) \gtrsim O(10^{-10})
\end{equation}
Since these estimates give only an order of magnitude for the 
lower and upper limits on corresponding branching ratios, we believe
that the region of $Br(o-Ps\to invisible) \simeq 10^{-9}-10^{-8}$ is 
of great interest for an observation of the effect of extra dimensions.

%%%%%%%%%%%%%%%%%%%%%%%%%%%%%%%%%%%%%%%%%%%%%%%

\section{Orthopositronium decay rate puzzle}
\label{sec:opuzzle}
Measurements of the positronium decays have a long history.
The measurements of the o-Ps lifetime performed after 1987 are summarized
in Table~\ref{results}.
Three precision measurements\cite{GAS87,Michigan_1989,Michigan_1990} 
of the o-Ps decay rate were originally performed at Ann Arbor, 
which reported decay rate values much larger, i.e., 5.1 -- 8.9 experimental 
standard deviations, than the QED prediction\cite{ADKINS-4}.
This discrepancy has been referred to as `orthopositronium lifetime puzzle', 
and was a long-standing problem. Measurements performed in Tokyo\cite{Tokyo_1995,Tokyo_2000,Tokyo_2003} and
the recent Ann Arbor measurement\cite{Michigan_2003} are in agreement with theory.
Experiments differed substantially in the technique
of formation of o-Ps.

\begin{table}[tbh]
\caption{Experimental results and theory (from Sillou\protect\cite{sillou}).}
{\begin{tabular}{@{}ccccccc@{}} \hline
Year & group/ref & Rate   & errors & Technique&exp - th & exp - th\\
&  & $\mu s ^{-1}$ &  (ppm) && & (sigma)\\ \hline
1987 & \hphantom{0}Ann Arbor [\cite{GAS87}] &   7.0516   \hphantom{000} &180& Gas&.01162& 8.9 \\
1989 & \hphantom{0}Ann Arbor [\cite{Michigan_1989}] &   7.0514    \hphantom{000} &200& Gas&.01142&8.2 \\
1990 & \hphantom{0}Ann Arbor[\cite{Michigan_1990}] &   7.0482    \hphantom{000} &230& Vacuum&.00822&5.1 \\
1995 & \hphantom{0}Tokyo [\cite{Tokyo_1995}]       &   7.0398    \hphantom{000} &412& Powder&-.00018&-.06 \\
2000 & \hphantom{0}Tokyo [\cite{Tokyo_2000}]       &   7.0399    \hphantom{000} &412& Powder &-.00008&0.0\\
2003 & \hphantom{0}Ann Arbor [\cite{Michigan_2003}] &   7.0404    \hphantom{000} &185& Vacuum&.00042 &0.32\\
2003 & \hphantom{0}Tokyo [\cite{Tokyo_2003}]       &   7.0396    \hphantom{000} &227&Powder &-.00038&.024\\
2000 & \hphantom{0}AFS[\cite{ADKINS-4}]      &  7.039979\hphantom{000} &1.6& Theory& &\\ \hline
\label{results}
\end{tabular}}
\end{table}

To elucidate discrepancies, a variety of experiments have 
been carried out to search for the exotic 
decay mode of o-Ps, resulting in no evidence\cite{EXOTIC-LL,EXOTIC-SL,EXOTIC-IV,EXOTIC-UB,EXOTIC-TW,EXOTIC-FOUR,Atoian:tz}\@.
Some of these searches are reported in Table~\ref{tab:exotic}.
Note that  best present limit for the branching ratio for invisible channels of $o-Ps$ is\cite{EXOTIC-IV} 
\begin{equation}
 Br(o-Ps \rightarrow invisible ) < 2.8 \times 10^{-6}
\end{equation}
However, this result was obtained in powder, and hence some of the comments
expressed in Section~\ref{sec:mirrorworld} might be relevant\cite{Gninenko:dr}.
As long as $t \gg 1/\Gamma_{coll}$, 
we have
\begin{equation}
\Gamma^{obs} \simeq\Gamma_{3\gamma}\left(1 + {2\omega^2 \over \Gamma_{coll}\Gamma_{3\gamma}}\right). 
\end{equation}
The difference between the higher decay rate measured in the vacuum
cavity experiment,
relative to the value predicted by theory,  can be expressed as
\begin{equation}
\Gamma_{exp}- \Gamma_{3\gamma} \simeq {2\omega^2 \over \Gamma_{coll}\Gamma_{3\gamma}} 
\end{equation} 
For  $\Gamma_{coll} \lesssim 3\Gamma_{3\gamma}$,
which should be applicable to the Ann Arbor cavity experiment (see Table~\ref{tab:gammcoll}), one finds
that
\begin{equation}
\omega^2 \sim 2\times 10^{-3}\Gamma^2_{3\gamma} \Rightarrow
\epsilon \lesssim 10^{-6}. 
\label{ep}
\end{equation}
Thus, the limit of Eq.(\ref{ep}) is still not strong enough compared to 
the BBN (see Figure~\ref{fig:footsum}).

Note, that the Tokyo results\cite{Tokyo_1995,Tokyo_2000,Tokyo_2003} and 
measurements in vacuum\cite{Michigan_1990,Michigan_2003}  are still consistent 
with the hypothesis of the mirror matter. Indeed in the Tokyo experiment 
the effect is suppressed by the very large collision rate of o-Ps 
in the powder, so it is irrelevant. The contribution of the mirror 
matter effect to the o-Ps decay rate in vacuum is allowed to be at the 
level of less than $\Delta \Gamma_{mirror}\simeq100 ppm$ according to the 
BBN limit on $\epsilon$. This is consistent with an "exotic" 
contribution to the  o-Ps decay rate, which is allowed to be 
$\Delta \Gamma_{exotic} < \Gamma_{Michigan} - \Gamma_{QED} \simeq 400 ppm$, as one 
can derive  from the difference between the new 
Michigan result and AFS calculations, see Table~\ref{results}.

%In conclusion, the agreement
%with the Tokyo experiment can be explained because of the
%very large collision rate of the orthopositronium in the powder.
%However because of the two different collision rates of the 
%two Ann Arbour experiments, they cannot both be explained.
%If we ignore this gas result then
%the discrepancy between the theory/Tokyo results and the
%Ann Arbour vacuum cavity experiment can be explained by
%the orthopositronium-mirror orthopositronium oscillation
%mechanism.  This seems however to be in contraction with
%limits from BBN.

\begin{table}[tb]
\caption{Upper limits on the branching ratios of several exotic o-Ps decays.}
{\begin{tabular}{@{}cccc@{}} \hline
\hline
Decay Mode & 90$\%$ upper limit & Comments & Group\\
\hline
$\gamma+X$ & 1.1 ppm & $X$ long--lived boson & CERN, Moscow,\\
           &        & m$_{X}<$800keV                    & Tokyo, Heidelberg\\
\hline
$\gamma+X\to \gamma + 2\gamma $ & 400 ppm & Short--lived boson & Moscow,\\ 
                                &        & m$_{X}<$900keV     & Tokyo\\
\hline
$\gamma\gamma $ & 3.5 ppm & Forbidden by angular  &  Ann Arbor,\\ 
                &         & momentum conservation &  Tokyo\\
\hline
$\gamma\gamma\gamma\gamma $ & 2.6 ppm & Forbidden by C--parity &  \\
\hline
$\gamma+X_1+X_2$ & 44 ppm & m$_{X_1}$+m$_{X_2}<$900keV & ETHZ- Moscow\\
\hline
Invisible & 2.8 ppm & Not in vacuum & Moscow, Tokyo\\
\hline
\end{tabular}}
\label{tab:exotic}
\end{table}

\subsection{Status of the QED calculations}
Theoretical aspects in theoretical predictions of quantities
related to positronium have been summarized at this workshop
by Karschenboim\cite{karschenboim}, Penin\cite{penin} and Smith\cite{smith}.

Positronium,  an electromagnetic bound state of the electron $e^-$ and
the positron $e^+$, is the lightest known atom.
Thanks to the smallness of the electron mass $m_e$ the strong and
weak interaction effects are negligible and its properties can be calculated
perturbatively in quantum electrodynamics (QED) as an expansion in
Sommerfeld's fine structure constant $\alpha$ with very high precision only
limited by the complexity of the calculations.
Positronium is thus a unique laboratory for testing the QED theory of 
bound systems.   The theoretical analysis is, however, complicated 
in comparison to other hydrogen-like atoms by
annihilation and recoil effects.
At the same time due to  negligible 
short-distance effects of the virtual 
strongly  interacting   heavy  particles,
positronium could be a sensitive probe
of the ``new physics'' at long distance.

The present theoretical knowledge of the 
decay rates (widths) of the $^3S_1$ orthoposit\-ronium (o-Ps)
and $^1S_0$ parapositronium (p-Ps) 
ground states 
to two and three photons, respectively, may be summarized as follows\cite{penin}:
\begin{eqnarray}
\Gamma_o^{\rm th}&=&{2(\pi^2-9)\alpha^6m_e\over 9\pi}
\left\{1+{\alpha\over\pi}10.286\,606 (10)+
\left({\alpha\over\pi}\right)^2\left[
{\pi^2\over 3}\ln{\alpha}
+44.87(26)\right]\right.
\nonumber\\
&&\left.
+{\alpha^3\over\pi}\left[-{3\over2}\ln^2{\alpha} 
+\left(3.428\,869(3)-{229\over 30}-8\ln2\right)\ln{\alpha}
+{D_o\over \pi^2}\right]\right\}\,,
\label{sort}
\\
\Gamma_p^{\rm th}&=&{\alpha^5m_e\over 2}
\left\{1+{\alpha\over\pi}\left({\pi^2\over 4}-5\right)
+
\left({\alpha\over\pi}\right)^2\left[
-2\pi^2\ln{\alpha}
+5.1243(33)\right]\right.
\nonumber\\
&&\left.
+{\alpha^3\over\pi}\left[-{3\over2}\ln^2{\alpha} 
+\left({533\over 90}-{\pi^2\over 2}+10\ln{2}\right)\ln{\alpha}
+{D_p\over \pi^2}\right]\right\}\,.
\label{spar}
\end{eqnarray}
The coefficients $D_{o,p}$ parameterize the unknown nonlogarithmic
${\cal O}(\alpha^3)$ terms.  
The calculation of missing { ${\cal O}(\alpha^3)$} 
nonlogarithmic terms would be one of the most complicated 
perturbative calculations in quantum field theory though
conceptually the problem is clear, all the necessary
tools are at hand and a number of partial results have been obtained.
Currently the 
experimental uncertainty exceeds the theoretical 
one by two orders of magnitude for positronium decay rates and by a 
factor of two for HFS. A lifetime experiment with an accuracy better than
50~ppm would be able to test the second-order correction which is
$\simeq 45\left({\alpha\over\pi}\right)^2\approx 200\ ppm$.
New measurements of much  higher
accuracy  would be mandatory to unambiguously 
confirm or confront the   QED predictions to higher-levels
and  to inspire the theorists to complete the
${\cal O}(\alpha^3)$  computations.

\subsection{Orthopositronium decay rate measurements}
As some fraction of o-Ps inevitably results in `pick-off' annihilations 
due to collisions with atomic electrons of the target material, 
the observed o-Ps decay rate 
$\lambda_{obs}$ is a sum of the intrinsic o-Ps 
decay rate $\lambda_{{\rm o}\mbox{-}{\rm Ps}}$ and the pick-off 
annihilation rate into $2\gamma$'s, 
$\lambda_{pick}$, i.e.,
\begin{equation}
\lambda_{obs}(t)=\lambda_{3\gamma}+\lambda_{pick}(t).
\end{equation} 
$\lambda_{pick}(t)$ is proportional 
to the rate of o-Ps collisions with the target materials, i.e.; 
$\lambda_{pick}=n\sigma_a v(t)$, where $n$ is 
product of the density of the target, $\sigma_a$ the annihilation cross-section, 
and $v(t)$ the time dependent velocity of o-Ps. 
Due to the thermalization process of o-Ps, 
this necessitates expressing 
$\lambda_{pick}$ as a function of time whose properties are dependent on 
the surrounding materials. 
Thermalization process should be carefully treated even in the 
cavity experiment\cite{Michigan_1990}.
Although pickoff correction is small in cavities, 
disappearance of o-Ps through the cavity entrance aperture 
has large contribution to $\lambda_{obs}$.
This disappearance rate is also proportional to $v(t)$, as the same reason.
Since the rate of elastic collision is extremely small in cavities, 
it takes much time, longer than 1 $\mu s$,
to thermalize well, and the disappearance rate still depends 
strongly on time.
%$\lambda_{obs}(t)$ should be considered to depend on time
%even in the cavity experiment.

In the Ann Arbor measurements\cite{GAS87,Michigan_1989,Michigan_1990}, 
$\lambda_{obs}$'s were measured by varying the 
densities of the target materials, size of the cavities and
also the entrance aperture of the cavities.
The extrapolation to zero density or aperture was expected to 
yield the decay rate in a vacuum, $\lambda_{3\gamma}$, under 
the assumption of quick thermalization (shorter than 170-180 nsec)
with constant o-Ps velocity. 
However, this assumption contains a serious systematic error as pointed out in 
reference\cite{PEKIN,ASAI95,asai}\@.

\begin{figure}[t!]%
  \begin{center}
    \includegraphics[width=18pc,keepaspectratio]{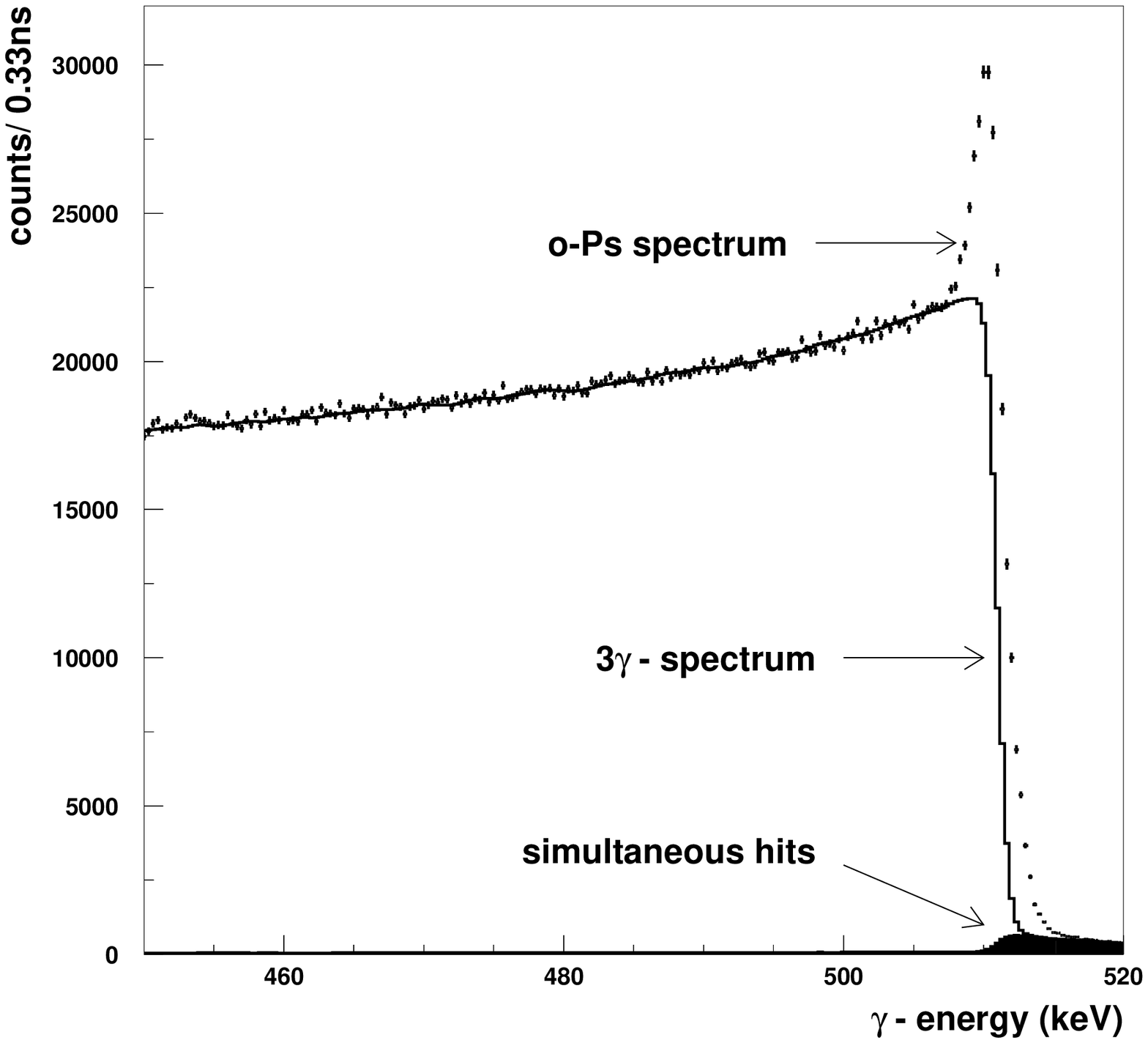}
    \includegraphics[width=18pc,keepaspectratio]{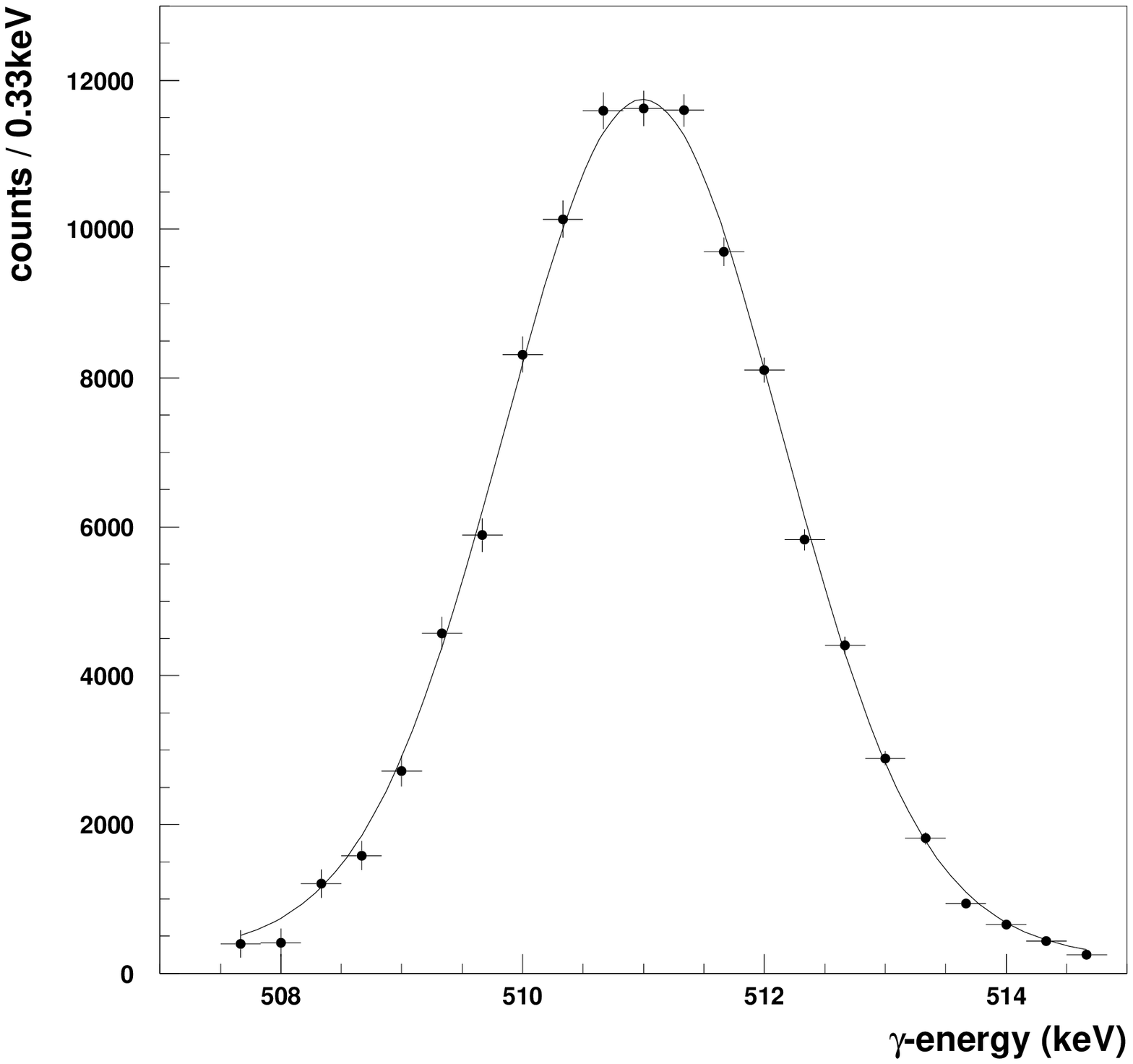}
  \end{center}
\caption{Tokyo experiment (Figures from Asai\protect\cite{asai}): (a) Energy spectrum of o-Ps decay $\gamma$'s obtained by Ge detectors. 
Dots represent data points in a time window of $150-700~ns$, and the solid 
line shows the $3\gamma$-decay spectrum calculated by Monte Carlo 
simulation. Shaded area indicate simultaneous hits estimated by the simulation. 
(b) Pick-off spectrum obtained after subtracting the $3\gamma$ contribution 
from the o-Ps spectrum. The solid line represents the fit result.}
\label{fig:ops-3gamma}\
\end{figure}

The Tokyo group proposed an entirely new method\cite{ASAI95}, 
which is free from above-mentioned systematic error.
The energy distribution of photons from the 3-body decay is 
continuous below the steep edge at 511~keV, 
whereas the pick-off annihilation is 2-body which produces 
a 511~keV monochromatic peak. 
Energy and timing information are simultaneously
measured with high-energy resolution germanium detectors such 
that $\lambda_{pick}(t)/\lambda_{3\gamma}$ can be determined
from the energy spectrum of the emitted photon. 
Once a precise thermalization function is obtained, $\lambda_{pick}(t)$ will 
contain all information about the process. 
The population of o-Ps at time $t$, 
$N(t)$ can be expressed as \begin{equation}
N(t)=N_0' \exp\left(-
\lambda_{3\gamma}\int^{t}_0\left(1+\frac{\lambda_{pick}(t')}
{\lambda_{3\gamma}}\right)dt'\right).
\end{equation}
Providing the ratio is determined as a function of time,
the intrinsic decay rate of o-Ps, $\lambda_{3\gamma}$, can be directly obtained
by fitting the observed time spectrum (See Figure~\ref{fig:ops-3gamma}).

Asai \textit{et al.} obtained the decay rate of $7.0398(29)$
and $7.0399(25)~\mu s^{-1}$ independently\cite{ASAI95,Tokyo_2000}, 
which are consistent with the non-relativistic QED calculation\cite{ADKINS-4}, 
and quite differ from the results originally obtained at Ann Arbor\cite{GAS87,Michigan_1989,Michigan_1990},
7.0482(16)--7.0516(13)$~\mu s^{-1}$\@. The 2003 Ann Arbor result is
$7.0404(13)~\mu s^{-1}$.

The observed $\lambda_{pick}(t)$ indicates that o-Ps thermalization
is slow and it is a serious systematic problem in all experiments using an extrapolation.   
In 1998, the Ann Arbor group recognized that the incomplete thermalization can have 
serious consequences in their measurements\cite{Michigan_1988_thermal}, but they did not 
yet update the orthopositronium lifetime. 

Regardless, still several issues existed in the first results of Asai \textit{et al.}\cite{asai}: 
(i) accuracy was $350~ppm$, being worse than those of the other
experiments\cite{GAS87,Michigan_1989,Michigan_1990}, (ii) there were unknown systematic uncertainties before
$t_{start}=200~ns$, and therefore to remove this uncertainty, 
final results were obtained using data after $220~ns$,  
and (iii) systematic error regarding the Stark effect was not estimated.
Improving results by considering these problems has been contributed
to these proceedings, where the final result using this new method were reported\cite{asai}.
The obtained decay rates are 
$\lambda_{3\gamma}=7.03991\pm0.0017(stat.)~\mu s^{-1}$ for RUN 1 and 
$7.03935\pm0.0017(stat.)~\mu s^{-1}$ for RUN 2, 
which are consistent with each other. 
It turns out that the thermalization process and pickoff ratio are
different within each run, however, excellent agreement between two runs is obtained. 
This should indicate that the method correctly takes into account thermalization and 
pickoff correction. 

%----------------------
%----------------------

%%%%%%%%%%%%%%%%%%%%%%%%%%%%%%%%%%%%%%%%%%%%%%%%%%%

\subsection{A personal summary}
Since the improvement of the  o-Ps decay rate in vacuum by the new calculations\cite{ADKINS-4}, 
the theoretical error is two orders of magnitude smaller than the experimental one. One could therefore
only agree with the conclusion from the latest AFS paper\cite{ADKINS-4}: 
``\textit{Éobviously, no conclusion can be drawn until the experimental situation is clarified}''.

\begin{figure}[htb!]
%\begin{center}
\vspace{-0cm}\hspace{-0.cm}{\epsfig{file=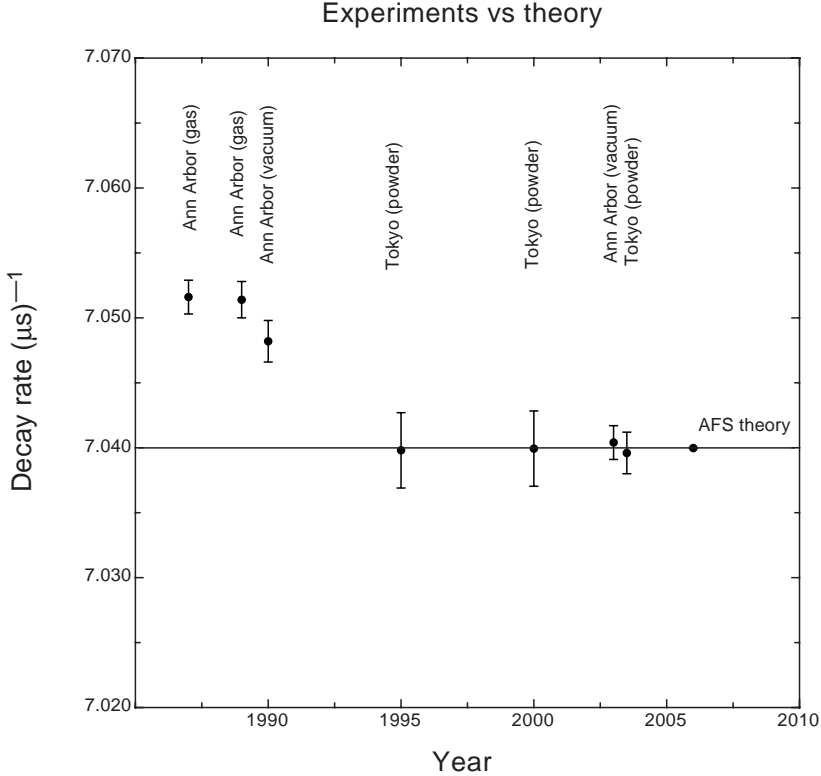,width=110mm}}
%\end{center}
\caption{\em Summary of the experimental results on o-Ps-lifetime measurements as a function
of the year. The last point is the theoretical value from AFS\protect\cite{ADKINS-4}.}
\label{fig:summgraph}
\end{figure}

The current experimental situation concerning the measurements of the orthopositronium
decay rate is graphically summarized in Figure~\ref{fig:summgraph}, where the decay rate
is plotted as a function of the year of the measurement. The last point is the theoretical value.
Different experiments have different systematic problems.

The much excitement that arose when the measurements performed by the Michigan group first 
in gas then in vacuum did not agree with theory seemed to have vanished
since the Michigan group has published a new result
which is now in agreement with the theoretical value, somewhat in contradiction
with their earlier results. The easy shortcut is to assume
that the o-Ps-lifetime puzzle is closed. Should we however indeed consider
that the o-Ps lifetime puzzle is solved and hence assume that 
further experimental investigations with positronium are irrelevant?

\begin{figure}[htb!]
%\begin{center}
\vspace{-0cm}\hspace{-0.cm}{\epsfig{file=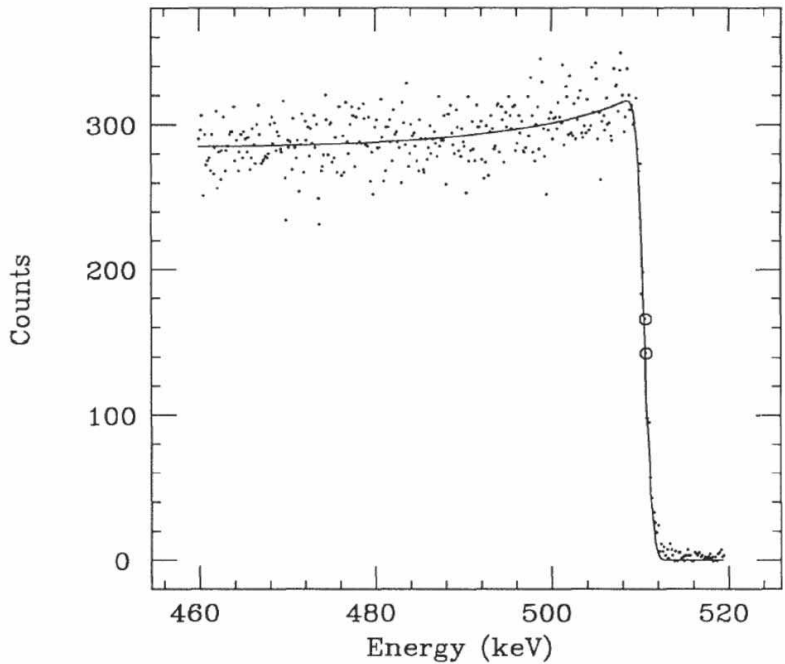,width=110mm}}
%\end{center}
\caption{\em Ann Arbor experiment (from Gidley et al.\protect\cite{Michigan_1991_2gam}): Energy
spectrum near 511 KeV measured with a precision Ge detector. No evidence for $2\gamma$ events (pickoff
signal) is visible.}
\label{fig:mich_2gammas}
\end{figure}

In the following, we play the Devil's advocate and 
list a few comments relative to the Ann Arbor experiments:
\begin{itemize}
\item The Michigan group concludes that they did not take properly into
account the ``backscattered'' high energy positronium component.
\item This interpretation seems to be in contradiction with the results of
an experiment performed with the same setup to search for a $2\gamma$
decay channel of o-Ps\cite{Michigan_1991_2gam}. The energy
spectrum near 511 keV measured with a precision Ge detector is shown
in Figure~\ref{fig:mich_2gammas}. No evidence for $2\gamma$ events (pickoff
signal) is visible. They exclude these decays at the level of $\simeq 200\ \rm ppm$, which
is a factor 5 smaller than the original discrepancy.
\end{itemize}
Similarly, a certain number of comments can be addressed to the Tokyo experiments:
\begin{itemize}
\item Systematic errors give a substantial contribution to the measurement error. In particular,
the correction due to pickoff in the high density target is large\cite{asai}:
\begin{equation}
\lambda_{pick}(t>200ns)\approx 0.01\times \lambda_{3\gamma} = (10^4 ppm)\times \lambda_{3\gamma}
\end{equation}
while the original discrepancy was at the level of 1000 ppm. Hence, the correction is an order of magnitude
larger than the original effect.
\item The experimental data seemed to be too consistent with theory. Indeed, the difference
between experiment and theory in sigmas are resp. $-0.06$, $0.0$ and $0.024$ for
the resp. 1995, 2000 and 2003 measurements. The statistical probability to obtain
three measurements within $0.06\sigma$ is $\approx 10^{-4}$. The origin of this could be related
to the fact that measurements are systematic dominated. Clearly, many of these errors
are correlated and as a result the various measurements cannot be easily combined.
\end{itemize}

The naive outcome from this non-exhaustive list is that new experiments to measure
the orthopositronium decay rate in vacuum in which the pickoff rate can be measured
directly via $2\gamma$ decays would bring new valuable information.

In addition, the 
current level of the theoretical precision
is about  two orders of magnitude better than the 
experimental one. Thus,   further 
positron beam based  experiments to measure the
$o-Ps$ decay rate in vacuum are required and are of great  interest
to test high-order QED corrections.
In general, the tendency is to go towards
high granularity hermetic detector in order to also discriminate on event topologies.
This allows simultaneous fit of time distributions in conjunction with 
simple topological cuts.

In parallel, other experimental approaches like the direct search for invisible decays
of o-Ps should be pursued, as discussed in Section~\ref{sec:invis}.

%%%%%%%%%%%%%%%%%%%%%%%%%%%%%%%%%%%%%%%%%%%%%%%%%%%%
%%%%%%%%%%%%%%%%%%%%%%%%%%%%%%%%%%%%%%%%%%%%%%%%%%%%
%%%%%%%%%%%%%%%%%%%%%%%%%%%%%%%%%%%%%%%%%%%%%%%%%%%%

\section{Positronium spectroscopy -- hyperfine splitting}

\begin{figure}[htbp] 
\centerline{\psfig{file=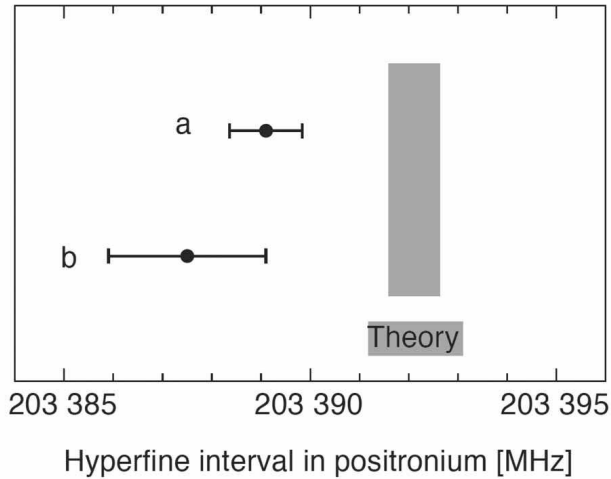,width=9cm}}
\caption{The positronium $1S$ hyperfine splitting: a comparison of theory to
  experiments. Figure from Ref.\protect\cite{karschenboim}\label{f:HFS}} 
\end{figure}

Positronium HFS,
$\Delta\nu=E\left(1^3S_1\right)-E\left(1^1S_0\right)$,
where $E\left(1^1S_0\right)$ and $E\left(1^3S_1\right)$ are the energy levels
of p-Ps and o-Ps ground state,  is the most precisely measured
quantity in positronium spectroscopy as far as the absolute precision is
concerned however measurements are not recent\cite{Mil,Rit}.
On the  theoretical side we have
\begin{eqnarray}
\Delta\nu^{\rm th}&=&{7m_e\alpha^4\over 12}\left\{1
-\frac{\alpha}{\pi}\left(\frac{32}{21}+\frac{6}{7}\ln2\right)
+\left(\frac{\alpha}{\pi}\right)^2
\left[-\frac{5}{14}\pi^2\ln{\alpha}+\frac{1367}{378}-\frac{5197}{2016}\pi^2
\right.\right.
\nonumber\\
&&
\left.+\left(\frac{6}{7}+\frac{221}{84}\pi^2\right)\ln2
-\frac{159}{56}\zeta(3)\right]
+\frac{\alpha^3}{\pi}\left[-\frac{3}{2}\ln^2{\alpha}
-\left(\frac{62}{15}-\frac{68}{7}\ln2\right)\ln{\alpha}\right.
\nonumber\\
&&
\left.\left.
+{D_\nu\over \pi^2}
\right]\right\}\,.
\label{shfs}
\end{eqnarray}
The theoretical result exceeds 
by approximately 2.6 and 3.5
experimental standard deviations (see Figure~\ref{f:HFS}). 
On the other hand, the measured $1S-2S$ interval seems to be in agreement
with theory. This shows that
the understanding of the positronium spectroscopy is not satisfactory.

Hence, new experiments on spectroscopy of positronium 
would bring new valuable information.
These experiments are however very different than those to measure
the lifetime or search for invisible decay.
There are midway between particle and atomic physics experiment and
require in addition expertise in the field of lasers and optics. We are not
aware of any plans for such an experiment.

%\section{Test of fundamental symmetries}
%Test of basic symmetries of QED (Krasnikov, Felcini)
%Measure angular asymmetry of photons relative to Ps spin, polarized with ext. B-field
%Experimental results at level 10Ð3, could be improved to 10Ð5 (Felcini)
%What about measurement of photon polarisation (Òasymmetry in polarisationÓ, Okun\&Telegdi) ? How to measure it?
%Unique for C-violation (cannot be done with atomic physics)
%Competitive with atomic physics? Antihydrogen?
%Phenomenological, no model which could predict large violations in o-Ps
%Anomalous magnetic moment of Ps?

\section{New searches for invisible decay of positronium}
\label{sec:invis}

As already mentioned in the previous sections,
the new models that are relevant to the $o-Ps\to invisible$ decay mode 
predict (see Ref.\cite{Gninenko:jn} for a review) the existence either of i) extra-dimensions, or ii) fractionally charged particles, or iii) a new light vector gauge boson, or
 iv) dark matter of the mirror matter type. The required sensitivity in the 
branching ratio $Br(o-Ps\to invisible)$  for the possible observation of these phenomena has to be at least as low as $10^{-8}$.

\subsection{ETHZ-INR experiment in silica aerogel}

A new experiment, aimed at developing a $4\pi$ hermetic calorimeter with an extremely low
photon detection inefficiency, was presented by Crivelli at this workshop\cite{crivelli}.
The main components of the detector are: 
the positron source (${}^{22}$Na), the positron tagging system, composed of a scintillating fiber viewed by two photomultipliers (PM), the positronium formation $SiO_2$  target and a  hermetic $\gamma$--detector.
 The coincidence of the PM signals from the positrons crossing the fiber, opens the gate for the data acquisition (DAQ). In the off--line analysis the 1.27 MeV photon, which is emitted from the source simultaneously with the positron, is required to be in the trigger BGO counter 
 resulting in a high confidence level of positron appearance in the positronium formation region. 
 A positron, which enters the $SiO_2$ target may capture an electron creating positronium. The calorimeter detects, either the direct $2\gamma$ annihilation in flight or the 2(3) photons from the para (ortho)--positronium decays in the target.  
The occurence of the $o-Ps \to invisible$ decay would appear as an excess of events with deposition in the calorimeter compatible
with zero above those expected from the Monte Carlo prediction or from the direct background measurement. 
This measurement presents a new feature of this type of experiment. The idea is to obtain a pure o--Ps decay energy spectrum by comparing two different spectra from the same target filled either with $N_2$ (low o-- Ps quenching rate) or with air, where the presence of  paramagnetic $O_2$ will quench the fraction of o--Ps in the target from 10$\%$ down to 3$\%$, due to the spin exchange mechanism: 
\begin{equation}
o-Ps+O( \uparrow \uparrow) \to p-Ps + O( \downarrow \uparrow)
\end{equation}
 Thus the subtraction of these properly normalized spectra will result in a pure o--Ps annihilation energy spectrum in the $\gamma$--detector.

\begin{figure}[htb]
\mbox{\epsfig{figure=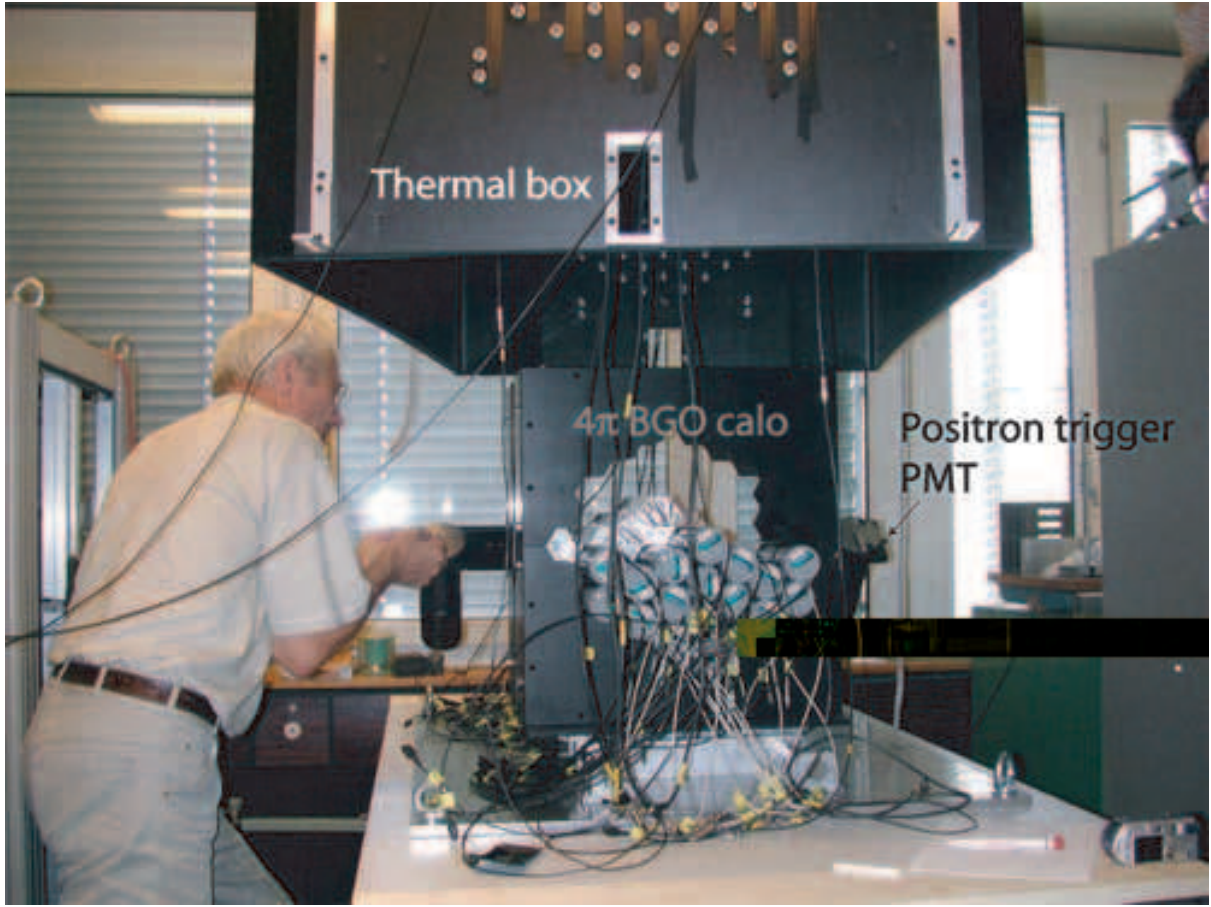,width=1.0\textwidth}}
\caption{\em ETHZ-INR experiment for $o-Ps\rightarrow invisible$ decay search in silica aerogel.}
\label{calophoto}
\end{figure}

The experiment is currently in commissioning phase (see Figure~\ref{calophoto}). Results are expected in 2004.

\subsection{ETHZ-INR experiment in vacuum}
\begin{figure}[htb!]
%\begin{center}
\vspace{-0cm}\hspace{-0.cm}{\epsfig{file=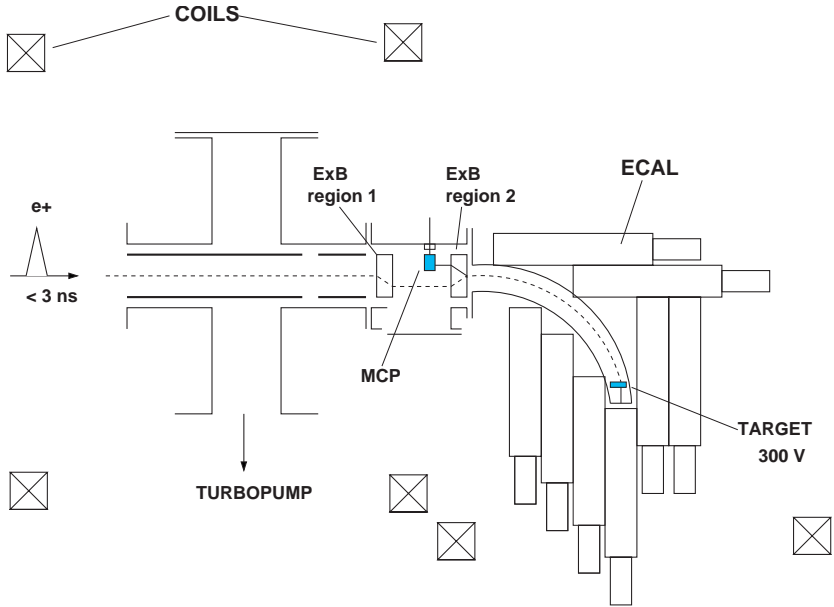,width=110mm}}
%\end{center}
\caption{\em Schematic diagram of the ETHZ-INR experimental setup.}
\label{tag}
\end{figure}

The concepts for a new experiment, designed with the goal to observe 
the  $o-Ps\to invisible$ decays, if its branching ratio
is greater than $10^{-7}$, were presented by Gninenko at this workshop\cite{gninenkoworkshop}.
Figure \ref{tag} shows a schematic view of the experimental setup.

Accordingly, the apparatus
is designed with several distinct parts:
i) a pulsed slow positron beam and a low-mass  target 
for efficient orthopositronium production in a vacuum cavity, 
ii) a  positron appearance tagging 
system with a high signal-to-noise ratio  based on a high performance MCP,
iii) an almost 4$\pi$  BGO crystal calorimeter (ECAL) surrounding the 
vacuum cavity for efficient detection of  annihilation photons. The 
 cavity has  as little wall mass as possible to minimize photon energy 
absorption. 

The occurrence of the 
$o-Ps\to o-Ps' \to invisible$ conversion would appear as an excess
of events with energy deposition comparable with zero  in the 
 calorimeter above those expected  from the
prediction of the background.
In case of a signal observation 
the number of excess events could be cross-checked by 
 small variations of experimental conditions  which affect the 
$o-Ps\to o-Ps'$ transition rate but do not result in a
 loss of energy from ordinary positron annihilations. 
The identification of signal events 
 relies on a high-efficiency measurement of the energy deposition from the 
annihilation of positrons. 

To achieve a sensitivity in the branching ratio of $10^{-7}$ 
in a reasonable amount of data-taking time, the rate of 
$o-Ps$ decays per second
 has to be as high as possible consistent with minimal reduction of the
  $o-Ps\to invisible$ signal efficiency and acceptably 
small dead time. For the  pulsed positron beam
design presented  by Gninenko at this workshop\cite{gninenkoworkshop},
 the trigger rate in the photon detector is expected to be 
$\simeq$ 100 Hz which is low enough to allow these 
events to be recorded without losses, and is high enough to reach
the expected sensitivity (see below) in a reasonable time.

Positrons from the pulsed beam 
are stopped in  the MgO target and either form positronium, i.e. $o-Ps$ or
$p-Ps$, or annihilate promptly into $2\gamma$'s. 
The secondary electrons (SE) 
produced by the positrons hitting the target are accelerated
 by the voltage applied to the target relative the grounded transport
 tube.  Then they are transported by a magnetic field in the
backward direction relative to the positrons 
moving in spirals along the magnetic field lines 
 and deflected to a microchannel plate (MCP)
 by a $E\times B$ filter.

The trigger for  data acquisition 
is generated by a coincidence 
within $\pm$3 ns of a pulse from the MCP 
 and the signal from the 
pulsed beam, which is synchronized with the positron 
arrival time at the target. 

It is hoped that the beam can serve several different experiments. Thus, the final beam construction
 should  compromise several design goals which are summarized as follows:
\begin{itemize}
\item beam energy range from 100 eV to 1000 eV,
\item beam intensity of $ \simeq 10^4-10^5$ positrons per second,  
\item pulse duration at the target $\delta t_T < 3$ ns for an 
 initial pulse duration at the moderator $\delta t_M \simeq 300-400$ ns,
\item repetition rate 0.3-1.0 MHz,
\item high peak/noise ratio, (single) Gaussian shape of the pulse, 
\item beam spot size at the target  position is of the order of a few 
millimeters assuming 3-5 mm $^{22}$Na source diameter,
\end{itemize}

\begin{figure}[htb!]
\begin{center}
\hspace{-0.cm}{\epsfig{file=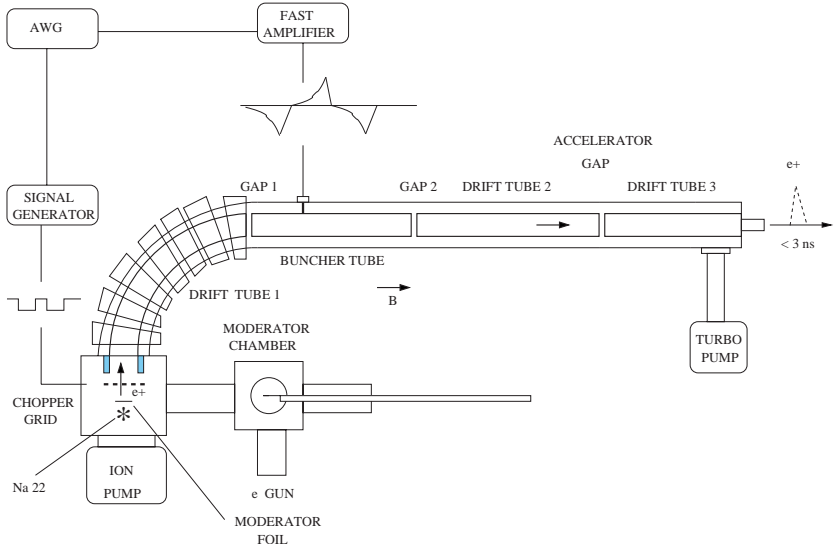,width=110mm}}
\end{center}
 \caption{\em ETHZ-INR experiment: Schematic illustration of the magnetically transported pulsed  
positron beam.}
\label{beam}
\end{figure}

Figure \ref{beam} shows schematic  illustration of the pulsed
positron beam design.
The positron pulsing section consists of a
chopper and a buncher and is  based on positron
velocity modulation combined with the RF bunching technique.
A positive potential ($\simeq 100$ eV) is applied to the moderator foil
 in order to insure the proper energy of the positrons at the buncher input.
Initial positron pulses with duration 300 ns are formed with the chopper
grid placed 2 mm apart from  the moderator foil. The pulsed voltage with 
an amplitude of about +5V applied to the chopper grid relative the moderator
foil will  stop slow positrons with energy  about 3 eV emitted from the
moderator. Fast positrons emitted from the source are
eliminated from the beam by the velocity analyzer (90 degrees curved
solenoid, placed downstream the chopper). When the voltage applied to the
chopper grid is zero, the positrons come through the chopper grid and
are accelerated in the gap between the chopper grid and first drift
tube (see Figure \ref{beam}).  Thus, positron pulses with a duration of 300 ns are
produced by this way. 

In the gap between the drift tube 1 and the buncher tube the
velocity of positrons from the 300 ns pulse is modulated by a nonlinear pulsed
voltage applied to the buncher tube relative to the drift tubes.
The buncher tube length is determined by a distance-of-flight of 
 positrons entering the buncher during 300 ns. In a second gap
between the buncher tube and a drift tube 2 the positron velocity is 
modulated  again by the same voltage pulse applied to the buncher.
   
The buncher voltage pulse is produced by an  arbitrary waveform generator
(AWG) connected to a fast post-amplifier whose shape for the
two-gap buncher is  determined appropriately. 
In this case
is determined by the ratio of the final and initial energy dispersion in the
positron beam pulse. Experimentally measured initial energy dispersion of
the moderated positrons is about 2 eV. 
Taking into account that the final
energy spread in the given two gap buncher is about 200 eV one  
expects a compression ratio of $\simeq$ 100.  

\begin{figure}[htb]
\mbox{\epsfig{figure=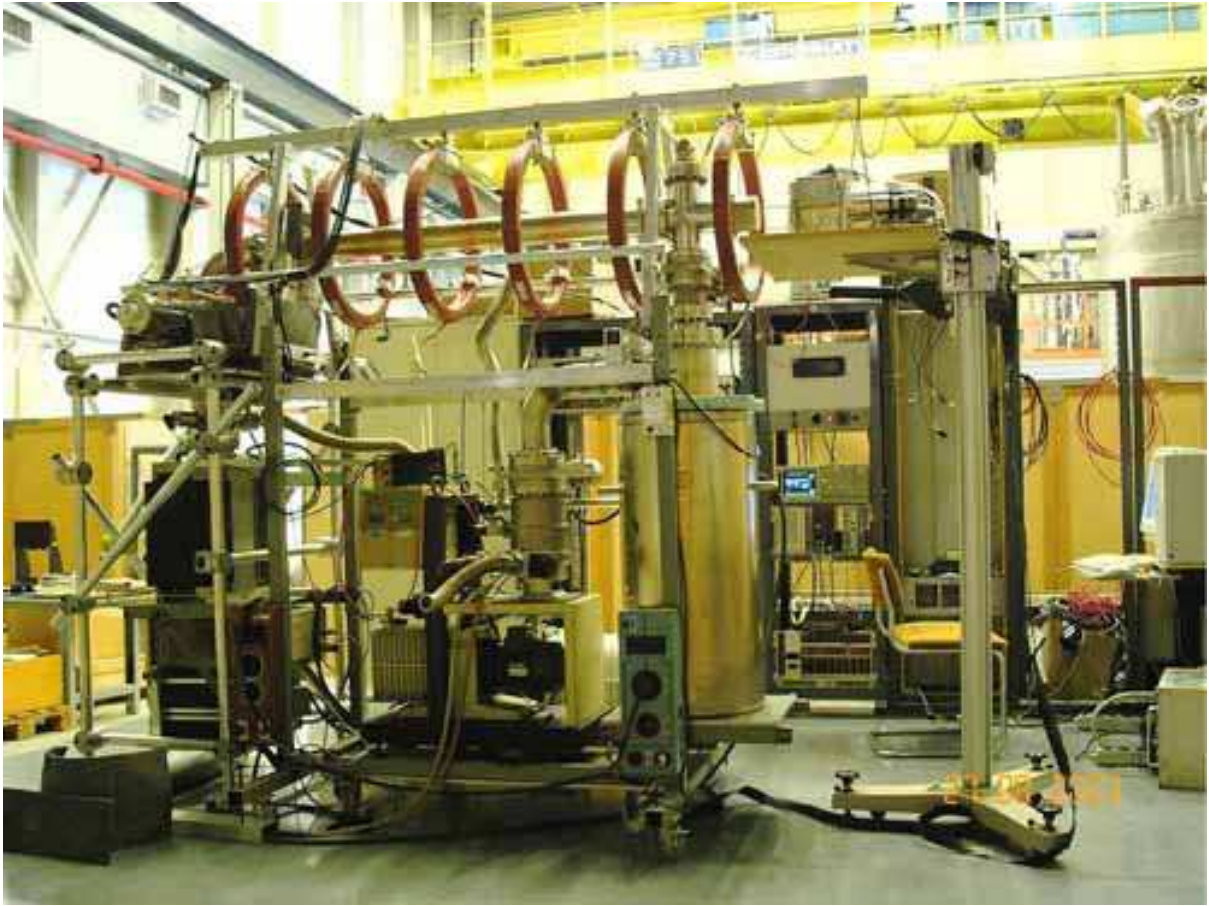,width=1.0\textwidth}}
\caption{\em ETHZ-INR setup: Positron beam (DC mode)
}
\label{fig:posit1}
\end{figure}

\begin{figure}[htb]
\mbox{\epsfig{figure=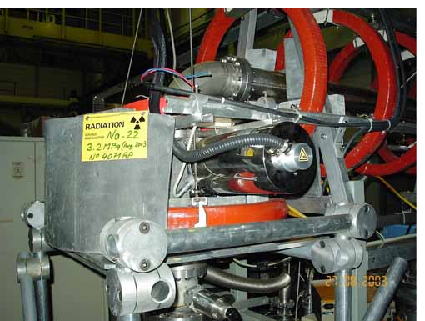,width=1.0\textwidth}}
\caption{\em ETHZ-INR setup: Na22 positron source and moderator chamber
}
\label{fig:posit2}
\end{figure}

The pulsed beam is currently under construction. In 2003 encouraging results in DC mode (See Figure~\ref{fig:posit1}) were
obtained confirming the correct design of the source, the moderator (see Figure~\ref{fig:posit2}) and the magnetic transport system.
Assembly of the pulsed beam is foreseen for 2004.

The experimental signature of  the $o-Ps\rightarrow invisible $ decay 
is an excess
 of events above the background at zero-energy deposition in the ECAL.
The 90\%-confidence level limit on the branching ratio for the 
$o-Ps \to invisible$ decay for a background free  experiment is given by 
\begin{equation}
S(90\%)=\frac{N(o-Ps\to invisible)}{N_{o-Ps} N_{coll}}
\label{sens}
\end{equation}
where $N(o-Ps\to invisible)=2.3$ and 
the terms in the denominator are the integrated number of produced
$o-Ps$'s ($N_{o-Ps}$), and the average number of $o-Ps$ collisions in the
cavity, respectively. The number
$N_{o-Ps}$  is defined as a product 
$N_{o-Ps}=R_{e^+}\cdot \epsilon_{o-Ps}\cdot \epsilon_{e^+}\cdot t $, 
where the first factor is the  number of 
delivered positrons per second on the target, the second one is the 
efficiency for 
$o-Ps$ production, and the third one is the efficiency of the 
secondary electron
 transportation from the target to the MCP in the 
positron tagging 
system. Taking $R_{e^+}= 2\times 10^3/sec$, $\epsilon_{o-Ps}= 20\%$ and
$\epsilon_{e^+}= 100\%$, 
 we expect $\approx 7\times 10^7$ prompt and $\approx 1.7\times
 10^7~o-Ps$ annihilations per day. Thus, $S(90\%)\simeq 10^{-7}$. 

    In case of the observation of zero-energy events, one of the 
 approaches would be to measure their number  
as a function of the residual gas pressure in the cavity.\
This would allow a good cross-check:
relatively small variations of gas pressure results in 
 larger peak variations at zero energy due to the damping  of
 $o-Ps\to o-Ps'$ oscillations. 

\subsection{Berkeley-LLNL experiment}
A collaboration at Lawrence Berkeley and Lawrence Livermore National
Laboratories is investigating the feasibility of extending the
sensitivity of such an experiment to the level of $10^{-9}$.  More information
was contributed by Vetter at this workshop\cite{vetter}. 
One
limitation is the counting time required to observe roughly
$10^{9}$ Ps formation triggers, while allowing a several microsecond
observation window to allow the o-Ps component to decay away such that
the exponential decay probability is less than the desired branching
ratio.  In such a case, the accidental rate should not give an
appreciable probability that a second decay would occur. 
Elementary calculations suggest that such a detector
would require a roughly 3 meter diameter in order to have negligible
probability for two 511 keV photons to each leave less than $\approx
100$~keV of energy in the detector volume.  The total mass of
such a volume of liquid scintillator would be roughly $10^{5}$ kg.
Such a geometry naturally addresses the hermeticity requirement, but
the sacrifice in compactness and larger mass compared to an array of
high-Z inorganic solid scintillators is clear.  
In such an experiment, the greatest challenge seems to be to provide
as clean a trigger as possible for the detector.  

\section{Conclusion}
From the large success of this workshop, it is clear that probing fundamental physics
with positronium is still an important issue. In this workshop, we have focused on
and discussed several new results and addressed new approaches to search for new
physics with positronium. Experiments are generally small but
difficult and many aspects of the current experiments are not fully understood. Further improved experiments
would be valuable information in different domains: 
\begin{enumerate}
\item in the context of the orthopositronium
lifetime, the theoretical precision is now almost two orders of magnitude better than the experimental
one. The first priority would be to test with high precision the theoretical prediction of
 he o-Ps decay rate computed to the second-order correction in $\alpha$.
A new generation experiment which could reach a precision
at the level of 50~ppm is under investigation. 
\item  in the context of spectroscopy of positronium, some problems still
need to be resolved, like for example the hyperfine splitting of the 1S level. 
\item In parallel to these
precision experiments, new experiments aimed at searching for invisible decays of the orthopositronium
are being designed or even constructed, pushing their sensitivity in the range of $10^{-7}$ for
the vacuum experiment and $10^{-8}$ for the experiment in aerogel. 
\end{enumerate}
These experiments will allow to further constrain models beyond the standard model. On the
other hand, finding such an effect would revolutionize our understanding of the physics of particles.

\section*{Acknowledgments}

I wish to express my warmest thanks to all participants of this workshop.
The help of collaboration 
in preparation of this paper and  many interesting discussions
with  P.~Crivelli, M. Felcini, S.~Gninenko and D. Sillou are greatly appreciated. 
Support from ETH Z\"urich, the Swiss National Science Foundation, INR, Moscow
and the participation of LMOPS Le Bourget du Lac, LAPP Annecy and CERI Orl\'eans
 is gratefully acknowledged.

%%%%%%%%%%%%%%%%%%%%%%%%%%%%%%%%%%%%%%%%%%%%%%%%%%%%%%%%%%%%
% Doing references:	                   	           %
%%%%%%%%%%%%%%%%%%%%%%%%%%%%%%%%%%%%%%%%%%%%%%%%%%%%%%%%%%%%


\begin{thebibliography}{99}

%%%%%%%%%%%%%%%%%%%%%%%%%%%%%%%%%%%%%%%%%%%%%%%%%%%%%%%%%%%%
% Command and Example    	                   	     %
%                                                          %
% \bibitem{REFERENCE_LABEL} AUTHORS NAMES,                 %
% {\it JOURNAL"S NAMES}{\bf VOLUME NUMBER}, PAGE (YEAR).   %
%                                                          %
% Three examples given below.	            	 	     %
%%%%%%%%%%%%%%%%%%%%%%%%%%%%%%%%%%%%%%%%%%%%%%%%%%%%%%%%%%%%

\bibitem{Deutsch} M. Deutsch, {\it Phys. Rev.} {\bf 82}, 455 (1951).

%\cite{Lee:qn}
\bibitem{Lee:qn}
T.~D.~Lee and C.~N.~Yang,
%``Question Of Parity Conservation In Weak Interactions,''
Phys.\ Rev.\  {\bf 104} (1956) 254.
%%CITATION = PHRVA,104,254;%%

%\cite{Okun:2001kg}
\bibitem{Okun:2001kg}
L.~B.~Okun,
%``Spacetime and vacuum as seen from Moscow,''
Int.\ J.\ Mod.\ Phys.\ A {\bf 17S1} (2002) 105
[arXiv:hep-ph/0112031].
%%CITATION = HEP-PH 0112031;%%

\bibitem{flv}   I. Kobzarev et al.,
Sov. J. Nucl. Phys. 3, 837 (1966); M. Pavsic, Int. J. Theor. Phys. 9,
229 (1974); R. Foot, H Lew and R. R. Volkas, Phys. Lett.
B272, 67 (1991).

\bibitem{berezhiani}
Z.~Berezhiani, to appear in Proceedings of
Workshop on Positronium Physics, Zurich (Switzerland), 30-31 May 2003.

\bibitem{foot}
R.~Foot,  to appear in Proceedings of
Workshop on Positronium Physics, Zurich (Switzerland), 30-31 May 2003.

\bibitem{hol}
B. Holdom, Phys. Lett. B166, 196 (1986).

\bibitem{s}
R. Foot, A. Yu. Ignatiev and R. R. Volkas,
Phys. Lett. B503, 355 (2001) [astro-ph/0011156].

\bibitem{gl}
S. L. Glashow, Phys. Lett. B167, 35 (1986).

%\cite{Foot:2000aj}
\bibitem{Foot:2000aj}
R.~Foot and S.~N.~Gninenko,
%``Can the mirror world explain the ortho-positronium lifetime puzzle?,''
Phys.\ Lett.\ B {\bf 480}, 171 (2000)
[arXiv:hep-ph/0003278].
%%CITATION = HEP-PH 0003278;%%

\bibitem{Michigan_1989} C.I. Westbrook, D.W. Gidley, R.S. Conti and A. Rich, {\it Phys. Rev.} {\bf A40}, 5489 (1989).

\bibitem{Michigan_1990} J.S. Nico, D.W. Gidley and A. Rich, {\it Phys. Rev. Lett.} {\bf 65}, 1344 (1990).

\bibitem{Tokyo_1995} S. Asai, S. Orito and N. Shinohara, {\it Phys. Lett.} {\bf  B 357}, 475-480 (1995).

\bibitem{f03}
R. Foot, hep-ph/0308254.

\bibitem{dama2}
R. Bernabei et al. (DAMA Collaboration), 
Riv. Nuovo Cimento. 26, 1 (2003) [astro-ph/0307403] and references
there-in.

\bibitem{BCV} Z. Berezhiani, D. Comelli and F.L. Villante,
{\it Phys. Lett.} \textbf{B503}, 362 (2001).

%\cite{Dubovsky:2000am}
\bibitem{Dubovsky:2000am}
S.~L.~Dubovsky, V.~A.~Rubakov and P.~G.~Tinyakov,
%``Brane world: Disappearing massive matter,''
Phys.\ Rev.\ D {\bf 62}, 105011 (2000)
[arXiv:hep-th/0006046].
%%CITATION = HEP-TH 0006046;%%

%\cite{Gninenko:2003nx}
\bibitem{Gninenko:2003nx}
S.~N.~Gninenko, N.~V.~Krasnikov and A.~Rubbia,
%``Extra dimensions and invisible decay of orthopositronium,''
Phys.\ Rev.\ D {\bf 67}, 075012 (2003)
[arXiv:hep-ph/0302205].
%%CITATION = HEP-PH 0302205;%%

\bibitem{GAS87}
{C. I. Westbrook, D. W. Gidley, R. S. Conti, and A. Rich}, {\it Phys. Rev. Lett}. {\bf 58}
  1328 (1987).

\bibitem{Tokyo_2000} 
%\cite{Jinnouchi:2000hg}
%\bibitem{Jinnouchi:2000hg}
O.~Jinnouchi, S.~Asai and T.~Kobayashi,
 %``Measurement of orthopositronium decay rate using SiO-2 powder:  Integration
%of thermalization into time spectrum fitting procedure,''
arXiv:hep-ex/0011011.
%%CITATION = HEP-EX 0011011;%%

\bibitem{Michigan_2003} R.S. Vallery, P.W. Zitzewitz and D.W. Gidley,  {\it Phys. Lett.} {\bf  90},     (2003).

\bibitem{Tokyo_2003} 
%\cite{Jinnouchi:2003hr}
%\bibitem{Jinnouchi:2003hr}
O.~Jinnouchi, S.~Asai and T.~Kobayashi,
%``Precision measurement of orthopositronium decay rate using SiO-2 powder,''
Phys.\ Lett.\ B {\bf 572} (2003) 117
[arXiv:hep-ex/0308030].
%%CITATION = HEP-EX 0308030;%%

\bibitem{ADKINS-4}
{G. S. Adkins, R. N. Fell, and J. Sapirstein}, {\it Phys. Rev. Lett.} {\bf 84} 5086 (2000) and
{\it Ann. Phys.} {\bf 295} 136 (2002).

\bibitem{sillou} D.~Sillou,  to appear in Proceedings of
Workshop on Positronium Physics, Zurich (Switzerland), 30-31 May 2003.

\bibitem{EXOTIC-LL}
{S. Asai, S. Orito, K. Yoshimura, and T. Haga}, {\it Phys. Rev. Lett.} {\bf 66}  2440 (1991); \\
{S. Orito, K. Yoshimura, T. Haga, M. Minowa, and M. Tsuchiaki},
{\it Phys. Rev. Lett.} {\bf 63} 597 (1989).

\bibitem{EXOTIC-SL}
{T. Maeno, M. Fujikawa, J. Kataoka, Y. Nishihara, S. Orito, K. Shigekuni, Y.
  Watanabe}, {\it Phys. Lett.} {\bf B351} 574 (1995); \\
{S. Asai, K. Shigekuni, T. Sanuki, and S. Orito}, {\it Phys. Lett.} {\bf B323} 90 (1994); \\
{M. Tsuchiaki, S. Orito, T. Yoshida, and M. Minowa}, {\it Phys. Lett.} {\bf B236} 81 (1990)
 81.

\bibitem{EXOTIC-IV}
{T. Mitsui, R. Fujimoto, Y. Ishisaki, Y. Ueda, Y. Yamazaki, S. Asai, and S.
  Orito}, {\it Phys. Rev. Lett.} {\bf 70} 2265 (1993).

\bibitem{EXOTIC-UB}
{T. Mitsui, K. Maki, S. Asai, Y. Ishisaki, R. Fujimoto, N. Muramoto, T. Sato,
  Y. Ueda, Y. Yamazaki and S. Orito}, {\it Euro. phys. Lett.} {\bf 33} 111 (1996); \\
{A. Badertscher, P. Crivelli, M. Felcini, S.N. Gninenko, N.A. Goloubev, P. Nedelec, 
J.P. Peigneux, V.Postoev, A. Rubbia and D. Sillou} {\it  Phys. Lett. } {\bf B542} 
29 (2002) 

\bibitem{EXOTIC-TW}
{S. Asai, S. Orito, T. Sanuki, M. Yasuda, and T. Yokoi}, {\it Phys. Rev. Lett.} {\bf 66}
1298 (1991); \\
{D. W. Gidley, J. S. Nico, and M. Skalsey}, {\it Phys. Rev. Lett.} {\bf 66} 1302 (1991).

\bibitem{EXOTIC-FOUR}
{K. Marko and A. Rich}, {\it Phys. Rev. Lett.} {\bf 33} 980 (1974).

%\cite{Atoian:tz}
\bibitem{Atoian:tz}
G.~S.~Atoian, S.~N.~Gninenko, V.~I.~Razin and Y.~V.~Ryabov,
%``A Search For Photonless Annihilation Of Orthopositronium,''
Phys.\ Lett.\ B {\bf 220} (1989) 317.
%%CITATION = PHLTA,B220,317;%%

%\cite{Gninenko:dr}
\bibitem{Gninenko:dr}
S.~N.~Gninenko,
%``Limit On 'Disappearance' Of Orthopositronium In Vacuum,''
Phys.\ Lett.\ B {\bf 326} (1994) 317.
%%CITATION = PHLTA,B326,317;%%.

\bibitem{karschenboim} S.G.~Karschenboim,  to appear in Proceedings of
Workshop on Positronium Physics, Zurich (Switzerland), 30-31 May 2003.
\bibitem{penin} A.~Penin, ibid.
\bibitem{smith} C.~Smith, ibid.

\bibitem{PEKIN}
{S. Asai , T. Hyodo, Y. Nagashima, T.B. Chang and S. Orito}, {\it Materials 
Science Forum,} {\bf 619} 175 (1995).

\bibitem{ASAI95}
{S.Asai} {`` New measurement of orthopositronium lifetime''}, Ph. D. 
thesis, University of Tokyo (1994);\\
{S. Asai, S. Orito, and N. Shinohara}, {\it Phys. Lett.} {\bf B357} 475 (1995).

\bibitem{asai} S. Asai, O. Jinnouchi and T. Kobayashi,  to appear in Proceedings of
Workshop on Positronium Physics, Zurich (Switzerland), 30-31 May 2003.

\bibitem{Michigan_1988_thermal} M. Skalsey, J. J. Engbrecht, R. K. Bithell, R. S. Vallery, and D. W. Gidley,  {\it Phys. Rev. Lett.} {\bf  80},   17  (1998).

\bibitem{Michigan_1991_2gam} D.W.Gidley, J.S. Nico and M. Skalsey, {\it Phys. Rev. Lett.} {\bf 66}, 1302-1305 (1991).

\bibitem{Mil}
A. P. Mills, Jr.\ and G. H. Bearman, 
{\it Phys.\ Rev.\ Lett.} {\bf34}, 246 (1975);
A. P. Mills, Jr.,
{\it Phys.\ Rev.} {\bf A27}, 262 (1983).

\bibitem{Rit}
M. W. Ritter, P. O. Egan, V. W. Hughes, and K. A. Woodle,
{\it Phys.\ Rev.} {\bf A30}, 1331 (1984).

%\cite{Gninenko:jn}
\bibitem{Gninenko:jn}
S.~N.~Gninenko, N.~V.~Krasnikov and A.~Rubbia,
%``Positronium Physics Beyond The Standard Model,''
Mod.\ Phys.\ Lett.\ A {\bf 17}, 1713 (2002).
%%CITATION = MPLAE,A17,1713;%%

\bibitem{czar}A.Czarnecki and S.G.Karshenboim, hep--ph/9911410:\\
 J. Govaerts and M.Van Caillie{\it Phys.Lett.}{\bf B391},451 (1996).

\bibitem{crivelli} P.~Crivelli,  to appear in Proceedings of
Workshop on Positronium Physics, Zurich (Switzerland), 30-31 May 2003.

\bibitem{gninenkoworkshop} S.~Gninenko, ibid.

\bibitem{vetter}
P.~Vetter, ibid.

%%%%%%%%%%%%
%%%%%%%%%%%%
%%%%%%%%%%%%
%%%%%%%%%%%%
%%%%%%%%%%%%
%

%\bibitem{sergei}
%S. N. Gninenko, Phys. Lett. B326, 317 (1994).

%

%

%\bibitem{D-thesis}
%{O.~Jinnouchi}, ``Study of bound state QED: precision measurement of the
%  orthopositronium decay rate'', Ph. D. thesis, University of Tokyo (2001);\\
%%\cite{Jinnouchi:2003hr}
%%\bibitem{Jinnouchi:2003hr}
%O.~Jinnouchi, S.~Asai and T.~Kobayashi,
%%``Precision measurement of orthopositronium decay rate using SiO-2 powder,''
%Phys.\ Lett.\ B {\bf 572} (2003) 117
%[arXiv:hep-ex/0308030].
%%%CITATION = HEP-EX 0308030;%%

%\bibitem{SIO2}
%{Y. Nagashima, T. Hyodo, K. Fujiwara, and A. Ichimura}, {\it J. Phys.} {\bf B31} 329
%(1998);\\
%{Y. Nagashima, M. Kakimoto, T. Hyodo, K. Fujiwara, A. Ichimura, T. Chang, J.
%  Deng, T. Akahane, T. Chiba, K. Suzuki, B. T. A. McKee, and A. T. Stewart},
%{\it Phys. Rev. }{\bf A 52} 258 (1995).

%
%\bibitem{oPs_newcchan} M.I. Dobrolioubov, S.N. Gninenko, A. Yu Ignatiev, V. A. Matveev   ,  {\it J. Mod. Phys.} {\bf  A8},   2859  (1993).\\
%M.Skalsey , {\it Mater. Science Forum  Trans Tech Pub. Switzerland.} {\bf  209},   255-257  (1997). \\
%S.N. Gninenko, N.V. Krasnikov and A. Rubbia, {\it Mod. Phys. Lett.} {\bf  A17}   1713 (2002).

%
%\bibitem{Michigan_1989_porousfilm} D.W. Gidley, W.E. Frieze, T.L. Dull, A. F. Yee, E.T. Ryan and H.-M. Ho ,  {\it Phys. Rev.} {\bf  B 60},   5157  (1999).

%\bibitem{Michigan_1995_filmspectrum} D.W. Gidley, D.N. McKinsey and P.W. Zitzewitz  ,  {\it J. Appl. Phys.} {\bf  78},   1406  (1995).

%\bibitem{nico}
%J. C.~Nico, D. W.~Gidley, A.~Rich, and P. W.~Zitzewitz, Phys.Rev.Lett.
%{\bf 65}, 1344 (1990); J. C.~Nico, Ph. D. thesis, Univ. of Michigan, 1991.

%




\end{thebibliography}
\end{document}